\documentstyle[prd,aps,epsf,psfrag]{revtex}

\makeatletter
  \renewcommand{\theequation}{%
     \thesection.\arabic{equation}}
  \@addtoreset{equation}{section}
\makeatother  

\newcommand{\be}{\begin{equation}}
\newcommand{\ee}{\end{equation}}
\newcommand{\bea}{\begin{eqnarray}}
\newcommand{\eea}{\end{eqnarray}}
\newcommand{\bi}{\begin{itemize}}
\newcommand{\ei}{\end{itemize}}
\newcommand{\ben}{\begin{enumerate}}
\newcommand{\een}{\end{enumerate}}
\newcommand{\bt}{\begin{tabbing}}
\newcommand{\et}{\end{tabbing}}

\renewcommand{\arraystretch}{1.03}

\title{
\begin{flushright}
{\normalsize UTHEP-491}\\ 
{\normalsize UTCCS-P-5}\\
\end{flushright}
A scaling study of the step scaling function 
\\
in SU(3) gauge theory
with improved gauge actions
}

\author{
S. Takeda$^1$,
S. Aoki$^1$, 
M. Fukugita$^3$,
K-I. Ishikawa$^5$, \\
N. Ishizuka$^{1,2}$,
Y. Iwasaki$^{1}$, 
K. Kanaya$^1$, 
T. Kaneko$^4$,\\
Y. Kuramashi$^{1,2}$, 
M. Okawa$^5$,
Y. Taniguchi$^{1,2}$, 
A. Ukawa$^{1,2}$,
T. Yoshi\'e$^{1,2}$ \\
 (CP-PACS Collaboration)
}

\address{
$^1$Graduate School of Pure and Applied Sciences, University of Tsukuba, 
    Tsukuba, 305-8571, Japan \\
$^2$Center for Computational Sciences, University of Tsukuba, 
    Tsukuba, 305-8577, Japan \\
$^3$Institute for Cosmic Ray Research, University of Tokyo, 
    Kashiwa 277-8582, Japan \\
$^4$High Energy Accelerator Research Organization
(KEK), Tsukuba, Ibaraki 305-0801, Japan \\
$^5$Department of Physics, Hiroshima University, Higashi-Hiroshima, 
    Hiroshima 739-8526, Japan 
}

\date{\today}

\begin{document}
  \draft
  \tightenlines
  \maketitle

 \begin{abstract}
We study the scaling behavior of the step scaling function for 
SU(3) gauge theory, employing the renormalization-group improved 
Iwasaki gauge action and the perturbatively improved L\"uscher-Weisz 
gauge action.
We confirm that the step scaling functions from the improved gauge actions 
agree with that previously obtained from the plaquette action
within errors
in the continuum limit at both weak and strong coupling regions. 
We also investigate how different choices of boundary
counter terms for the improved gauge actions affect the scaling behavior.
In the extrapolation to the continuum limit,
we observe that the cut off dependence becomes moderate for the Iwasaki action,
if a perturbative reduction of scaling violations is applied to 
the simulation results.
We also measure the low energy scale ratio with the Iwasaki action,
and confirm its universality.
 \end{abstract}

 \section{Introduction}
  \label{sec:Introduction}

The strong coupling constant is one of the fundamental parameters of QCD.
The current world average leads to
$\alpha_{\overline{\rm{MS}}}(m_Z)=0.1172(20)$\cite{hagiwara}.
Lattice QCD calculations have a potential ability
to determine the strong coupling constant from
an experimental input at low energy scales.
In practice, however, one must relate the 
high energy perturbative QCD scale to the low energy hadronic scale.
Alpha Collaboration proposed the Schr\"odinger functional(SF) scheme 
as a vehicle for this purpose\cite{sforiginal,sffermion}, 
and it has been successfully 
applied to lattice QCD in various aspects
\cite{coefficient,jansen,current,guagnelli}. 
One of the most recent results related to our study is 
the running coupling constant of
two massless flavor QCD reported in Ref. \cite{first,recent}

Recently CP-PACS and JLQCD Collaborations
have started a project for $N_{\rm f}=3$ QCD simulations
\cite{ishikawa,okawa,cswnonperturbative,ishikawa2003,kaneko2003}.
These simulations are essential to understand the
low energy QCD dynamics for the real world in which three light quarks 
exist.
One of the targets of the project is to evaluate the strong coupling constant 
$\alpha_{\overline{\rm{MS}}}$ in $N_{\rm{f}}=3$ QCD  
using the SF scheme.
In the project Iwasaki gauge action\cite{Iwasaki} is
employed to avoid the strong lattice artifacts of the plaquette gauge
action found in $N_{\rm f}=3$ simulations\cite{okawa}.

In a previous study \cite{takeda}, as our first step toward evaluation of 
$\alpha_{\overline{\rm{MS}}}$ for $N_{\rm{f}}=3$, 
O($a$) boundary improvement coefficients in the SF scheme
have been determined for various improved gauge actions
up to one-loop order in perturbation theory.
In addition the scaling violation in the step scaling function(SSF)
for the coupling 
have been analyzed perturbatively.
In the present paper, as the next step, 
we investigate the lattice cut off dependence of the SSF
non-perturbatively in quenched lattice QCD simulations with
improved gauge actions. 
The renormalization-group improved Iwasaki gauge action and 
the perturbatively improved L\"uscher-Weisz gauge action are employed.
We investigate the effect of various choices for
boundary improvement coefficients in detail, to find the best choice,
which will be used in our unquenched simulations in the future.
We also confirm the universality of the SSF and the low
energy scale ratio, by comparing our results with the previous ones obtained
by ALPHA Collaboration\cite{NPRquench,Necco}

The rest of this paper is organized as follows.
In Sec. \ref{sec:Preliminaries}, 
after a brief introduction of the SF scheme and 
its extension to improved gauge actions,
we specify the action and the O($a$) boundary improvement coefficients
used in our simulations. We then define 
the Schr\"odinger functional coupling constant,
the step scaling function and the low energy scale ratio.
In Sec. \ref{sec:Simulation},
we give details of simulations and present our results
with improved gauge actions for various choices for O($a$) improvement.
In Sec. \ref{sec:CutOff},
we investigate the lattice cut off dependence of
the step scaling function and the low energy scale ratio, 
and carefully take the continuum limit of these quantities,
in order to confirm their universality. 
Our conclusion is given in the last section, 
together with a discussion toward $N_{\rm f}=2$ and $3$ simulations.

 \section{Preliminaries}
  \label{sec:Preliminaries}

   \subsection{Schr\"odinger functional}
    \label{subsec:SF}

The SF scheme 
introduced by ALPHA Collaboration
is a powerful tool to
probe the energy evolution of physical quantities.
In the SF scheme, the theory is defined on a finite box of 
size $L^3\times T$ with the periodic boundary condition in 
the spatial directions and the Dirichlet boundary
condition in the time direction.
We set $T=L$ throughout this paper.
In the pure SU($3$) gauge theory with Wilson plaquette action $S[U]$,
the Schr\"odinger functional 
is given by
    \begin{equation}
      {\cal Z} 
      =
      \int D[U] e^{ - S[U] },
    \end{equation}
where the link variables $U(\mu,x)$ 
for the gauge fields satisfy the boundary conditions
    \begin{equation}
      \left.U(x,k)\right|_{x_0 = 0} 
      = 
      \exp \{ a C \} ,\mbox{\hspace{5mm}}
      \left.U(x,k)\right|_{x_0 = L} 
      = 
      \exp \{ a C^{\prime} \}.
     \label{eqn:CC}
    \end{equation}
Here $a$ is lattice spacing, and
$C$, $C^{\prime}$ are spatially constant diagonal matrices, which depend on 
the background field parameters $\eta$ and $\nu$ \cite{su3}.

An extension of the SF scheme to the improved gauge actions
was first discussed by Klassen \cite{klassen} in terms of 
a transfer matrix construction\cite{imptransfer}. 
In this formulation,
each boundary consists of two time slices, to achieve the tree-level 
O($a^2$) improvement.

In this paper, however, we adopt an alternative formulation\cite{aokiweisz},
which achieves the tree-level O($a$) improvement with only one time slice
at each boundary.
The dynamical variables to be integrated over are 
independent of the form of the 
action, whether plaquette or improved, and consist of
the spatial link variables $U(k,x)$ with $x_0 = a,\cdots,L-a$
and temporal link variables $U(0,x)$ with $x_0 = 0,\cdots,L-a$
on the cylinder with volume $L^3 \times L$.
This formulation is implemented more easily in numerical simulations.

   \subsection{Gauge action and O($a$) boundary improvement coefficients}
    \label{subsec:G-Oaimp}

The improved action we employ includes the plaquette and rectangle loops,
and is given by
    \begin{equation}
      S_{\rm{imp}}[U] 
      = 
      \frac{1}{g^2_0} 
      \sum_{{\cal C} \in {\cal S}_0} W_0({\cal C},g^2_0) 2{\cal L(C)}
    + \frac{1}{g^2_0} 
      \sum_{{\cal C} \in {\cal S}_1} W_1({\cal C},g^2_0) 2{\cal L(C)},
     \label{eqn:impaction}
    \end{equation}
with
    \begin{equation}
      {\cal L(C)} 
      = 
      \mbox{ReTr}[I - U({\cal C})],
    \end{equation}
where $W_i$ is a weight factor 
to be specified later and 
$U({\cal C})$ is an ordered product of 
the link variables along a loop ${\cal C}$
contained in a set ${\cal S}_0$(plaquette) or ${\cal S}_1$(rectangular).
${\cal S}_0$ and ${\cal S}_1$ consist of 
all loops of the given shape which can be 
drawn on the cylindrical lattice with the volume $L^3 \times L$.
The loops involve the ``dynamical links'' in the
sense specified above, and spatial links
on the boundaries at $x_0 = 0$ and $x_0 = L$.
In particular, rectangles protruding from the boundary
of the cylinder are not included.

\begin{figure}[]
   \begin{center}
       \psfragscanon
       \psfrag{time}[][][1.5]{time}
       \psfrag{space}[][][1.5]{space}
       \psfrag{T}[][][1.5]{$t=0$}
       \psfrag{P}[][][1.5]{$c_0 c^P_{\rm{t}}(g_0^2)$}
       \psfrag{R}[][][1.5]{$c_1 c^R_{\rm{t}}(g_0^2)$}
       \psfrag{RR}[][][1.5]{$c_1$}
       \psfrag{RT}[][][1.5]{$0$}
       \psfrag{PS}[][][1.5]{$c_0 c_s(g_0^2)$}
       \psfrag{RS}[][][1.5]{$0$}
       \scalebox{0.63}{\includegraphics{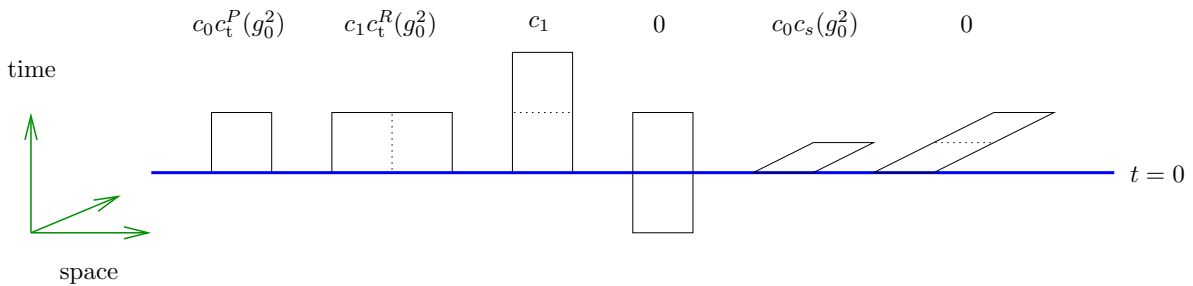}} 
   \end{center}
 \caption{The assignments of the weight factor for loops near 
the boundary $t=0$ eq. (\ref{eqn:W0}, \ref{eqn:W1}).
 \label{boundary}}
\end{figure}

One needs to choose the weight factors appropriately 
to achieve the one-loop level O($a$) improvement.
Among various possible choices, ours is given as follows.
    \begin{equation} 
      W_0({\cal C},g^2_0) 
      = 
      \left\{
       \begin{array}{ll}
         c_0 c_{\rm{s}}(g^2_0) 
       & \mbox{for } {\cal C} \in P_{\rm{s}} : 
          \mbox{Set of plaquettes that lie completely} 
       \\
       &
          \mbox{\hspace{19mm} on one of the boundaries, } 
       \\      
         c_0 c^P_{\rm{t}}(g^2_0) 
       & \mbox{for } {\cal C} \in P_{\rm{t}} : 
          \mbox{Set of plaquettes that just touch one} 
       \\
       &
          \mbox{\hspace{19mm} of the boundaries, } 
       \\      
         c_0        
       & \mbox{for } {\cal C} \in P_{\rm{other}} : 
          \mbox{otherwise, } 
       \end{array}
            \right.
     \label{eqn:W0}
    \end{equation} 
    \begin{equation} 
      W_1({\cal C},g^2_0) 
      = 
      \left\{
       \begin{array}{ll}
         0          
       & \mbox{for } {\cal C} \in R_{\rm{s}} : 
          \mbox{Set of rectangles that lie completely} 
       \\
       &
          \mbox{\hspace{19mm} on one of the boundaries, } 
       \\
         c_1 c^R_{\rm{t}}(g^2_0) 
       & \mbox{for } {\cal C} \in R^2_{\rm{t}} : 
          \mbox{Set of rectangles that have exactly two} 
       \\
       &
          \mbox{\hspace{19mm} links on a boundary, } 
       \\     
         c_1        
       & \mbox{for } {\cal C} \in R_{\rm{other}} : 
          \mbox{otherwise, }     
       \end{array}
            \right.
     \label{eqn:W1}
    \end{equation} 
with 
    \begin{eqnarray}
      c_0 c^P_{\rm{t}}(g^2_0)
      & = & 
      c_0 ( 1 + c^{P(1)}_{\rm{t}} g^2_0 + O(g^4_0) ), \label{eqn:ctP}\\
      c_1 c^R_{\rm{t}}(g^2_0)
      & = & 
      c_1 ( 3/2 + c^{R(1)}_{\rm{t}} g^2_0 + O(g^4_0) ),
\label{eqn:ctR}
   \end{eqnarray}
where the coefficients $c_0$ and $c_1$ of the improved gauge action are 
normalized such that $c_0 + 8 c_1 = 1$.
In this paper we consider
not only
the Iwasaki action ($c_1 = -0.331$, $c_2 = c_3 = 0$)\cite{Iwasaki}
but also
the L\"uscher-Weisz (LW) action ($c_1 = -1/12$, $c_2 = c_3 = 0$)\cite{LW}
for comparison,
since our perturbative analysis \cite{takeda} shows that
the LW action has a fairly small lattice artifact in the step scaling
function.
We call $c^P_{\rm{t}}(g^2_0)$ and $c^R_{\rm{t}}(g^2_0)$
O($a$) boundary improvement coefficients.
The assignments at the $t=0$ boundary are shown in Fig. \ref{boundary}.
The leading term of the O($a$) boundary improvement coefficients
in eqs.(\ref{eqn:ctP}) and(\ref{eqn:ctR})
can uniquely be determined from two requirements that
the tree-level O($a$) improvement is achieved and
the lattice background field satisfies the equation of motion at the 
boundaries\cite{aokiweisz}.
On the other hand, for the one-loop boundary terms,
we can freely set a relation between
$c^{P(1)}_{\rm{t}}$ and $c^{R(1)}_{\rm{t}}$,
since there is only one requirement for the one-loop O($a$) improvement.

Let us see how we specify the one-loop boundary terms.
In Ref. \cite{takeda},
one finds the following relation
to achieve the one-loop O($a$) improvement:
    \begin{equation}
      c_0 c^{P(1)}_{\rm{t}}
    + 4 c_1 c^{R(1)}_{\rm{t}}
      =
      A_1/2,
     \label{eqn:ct}
    \end{equation}
where $A_1$ is the coefficient of the $a/L$ term in the 
one-loop correction $m^{(0)}_1(L/a)$ to the SF coupling.
In our simulations we consider two choices: one called condition A is
given by 
    \begin{equation}
      c^{R(1)}_{\rm{t}}
      =
      2 c^{P(1)}_{\rm{t}},
     \label{eqn:conditionA}
    \end{equation}
and the other called condition B is specified by 
    \begin{equation}
      c^{R(1)}_{\rm{t}}
      =
      0.
     \label{eqn:conditionB}
    \end{equation}
The difference between the conditions A and B
is an O($a^5$) contribution in the one-loop correction to the
SF coupling\cite{takeda}.
Although this difference is tiny at one-loop order,
it may become larger at higher orders.
The values of one-loop boundary terms for each condition and $A_1$ are given 
in Table \ref{tab:ct}.
For the LW action, the difference between the two conditions is small, 
so we do not carry out simulations with the condition B.

   \subsection{Schr\"odinger functional coupling}
    \label{subsec:SFcoupling}

The SF with the improved gauge action is given by
    \begin{equation}
      {\cal Z}
      =
      e^{- \Gamma}
      =
      \int D[U] e^{ - S_{\rm imp}[U] },
     \label{eqn:SFimp}
    \end{equation}
where we impose the same boundary condition eq. (\ref{eqn:CC}) for       
the link variables as in the case of the Wilson plaquette action.
The SF coupling is defined through
the free energy $\Gamma$ in eq. (\ref{eqn:SFimp})
    \begin{equation}
      \bar{g}^2_{\rm{SF}}(L)
      =
      \left.
      k/\Gamma^{\prime} 
      \right|_{\eta=\nu=0}
      =
      k/
      \left.
      \left\langle
      \frac{\partial S}{\partial \eta}
      \right\rangle
      \right|_{\eta=\nu=0},
     \label{eqn:observable}
    \end{equation} 
where $k$ is a normalization constant
    \begin{eqnarray}   
      k 
      &=&
      12 \left(
           \frac{L}{a}
         \right)^2 
         \left[
           c_0 
           \left( 
             \sin 2 \gamma
           + \sin   \gamma
           \right)
           + 4 c_1 
           \left( 
             \sin 4 \gamma
           + \sin 2 \gamma
           \right)              
         \right],  
      \\
      \gamma 
      &=&
      \left(
        \frac{a}{L}
      \right)^2
      \left( 
        \eta + \frac{\pi}{3}
      \right).
    \end{eqnarray}
The renormalized coupling $\bar{g}^2_{\rm{SF}}(L)$
depends only on the scale determined by the box size $L$.
The $\eta$-derivative of the loop touching the boundaries 
(the left-most three loops in Fig. \ref{boundary}, for example),
contributes to the observable $\partial S / \partial \eta$
in eq. (\ref{eqn:observable}).

   \subsection{Step scaling function and Low energy scale ratio}
    \label{subsec:SSFandLmax}

The step scaling function (SSF) describes the
evolution of the renormalized coupling under
a finite rescaling factor $s$ (we take $s=2$ in the following)
     \begin{equation}
       \sigma(2, u) 
       =
       \left.
         \bar{g}^2(2L) 
       \right|_{\bar{g}^2(L) = u}.
     \end{equation}
By choosing the $n+1$-th initial value of $\sigma(2,u_{n+1})$ such that 
$u_{n+1} = \sigma(2,u_n)$, the non-perturbative evolution
of the running coupling can be constructed successively in order to
cover a wide range of the energy scale.

The SSF $\sigma(2,u)$ in the continuum theory is obtained by the
continuum limit of a lattice SSF $\Sigma(2,u,a/L)$
     \begin{equation}
       \sigma(2,u)
       =
       \lim_{a/L \rightarrow 0}
       \Sigma(2,u,a/L).
     \end{equation}
In this paper,
we study the SSF at a weak coupling $u=0.9944$
and a strong coupling $u=2.4484$,
where our results can be compared with
those of ALPHA Collaboration.

To fix the scale in a physical unit,
one needs to relate the box size $L$ prescribed at a certain value 
$\bar{g}^2_{\rm SF}(L)$ to some reference scale.
Following the conventional way,
we set $L=L_{\max}$ defined implicitly
     \begin{equation}
       \bar{g}^2_{\rm SF}(L_{\max})
       =
       3.480,
     \end{equation}
and adopt Sommer's scale $r_0$ \cite{Sommer}
as the reference scale.
Eventually this amounts to computing
the ratio $L_{\max}/r_0$ and extrapolating it 
to the continuum limit
  \begin{equation}
    L_{\rm{max}}/r_0
    =
    \lim_{a/{L_{\rm{max}}} \rightarrow 0}
    (L_{\rm{max}}/a)
    \times
    (a/r_0).
   \label{eqn:contlowenergy}
  \end{equation}

 \section{Simulation details and results}
  \label{sec:Simulation}

We follow the calculation procedure of Ref. \cite{su3}.
Simulations for the SSF on larger lattices are performed on CP-PACS
using 4 partitions of 64 PU's, while $r_0/a$ are calculated with 2 partitions of 
512 PU's.

As mentioned in Sec. \ref{subsec:SFcoupling},
the SF coupling is obtained by
calculating the observable $\partial S / \partial \eta$
for gauge configurations with the SF boundary condition.
When the number of spatial lattice points, $L/a$,
is a multiple of $4$,
the gauge configurations are generated by 
a combined five-hit pseudo-heat-bath algorithm
and an over-relaxation algorithm (HB).
The combination of one pseudo-heat-bath update sweep
followed by $N_{\rm OR}=L/2a$ over-relaxation sweeps
is called an iteration.
The measurement is implemented after each sweep,
i.e. $(1+L/2a)$ measurements are made per one iteration.
Because of a restriction of the HB method optimized for
the improved gauge action on CP-PACS,
we employ the hybrid Monte Carlo (HMC) algorithm
for $L/a$ being different from multiples of $4$, 
for instance $L/a=6$.
The step size for the molecular dynamics is adjusted
to achieve an acceptance rate in a range from $0.7$ to $0.8$.
The measurement is made for every trajectory.

Our computations for the renormalized coupling
$\bar{g}^2_{\rm SF}(L)$
are carried out on lattices $L/a=4$, $6$, $8$, $12$.
In this calculation, a re-weighting technique is used for a tuning of $\beta$
\cite{O3ssf}
such that $\bar{g}^2_{\rm SF}(L)$ 
becomes a certain prescribed value $u$ for each $L/a$.
And then,
using the same $\beta$,
a computation on a lattice with twice the linear size $2L/a$
gives $\bar{g}^2_{\rm SF}(2L)$.
The results are summarized in Table \ref{tab:weak} and \ref{tab:strong}
for the weak($u=0.9944$) and strong($u=2.4484$) couplings, respectively.
Errors in both $\bar{g}^2_{\rm SF}(L)$ and $\bar{g}^2_{\rm SF}(2L)$ 
are estimated by a jack-knife method.
The bin size for jack-knife errors is
$100$ iterations for the HB and $500$ trajectories for HMC, respectively.
The precision in $\bar{g}^2_{\rm SF}(L)$ is attained by accumulating around
$120 000-140 000$ iterations on the lattices $L/a=4$, $8$, $12$,
and $300 000$ trajectories on the lattices $L/a=6$.
As for $\bar{g}^2_{\rm SF}(2L)$, the number of iterations
is around $40 000-80 000$ to achieve their precision.
Errors in $\bar{g}^2_{\rm SF}(L)$
are propagated into $\Sigma(2,u,a/L)$, the lattice SSF,
where $u$ is the central value of $\bar{g}^2_{\rm SF}(L)$.
A formulae of the error propagation using 
a perturbative expansion of the SSF can be found in Ref. \cite{gehrmann}.

We performed an additional set of simulations with the Iwasaki action 
to determine the low energy scale. 
The tuning of $\beta$ to the conventional point 
$\bar{g}^2_{\rm SF}(L_{\max})=3.480$ 
and the error analysis
are made in the same way as mentioned above.
In Table \ref{tab:tuningbeta} we list the results,
which will be used in Sec. \ref{subsec:lowenergyscales}
as the first factor on the right hand side of 
eq. (\ref{eqn:contlowenergy}).
To complete the scale determination,
one needs the second factor in eq. (\ref{eqn:contlowenergy}).
In addition to the previous results of $r_0/a$ 
\cite{okamoto,topology,nf2},
we carried out simulations at $\beta=3.00$ and $3.53$
to cover the range of $\beta$ in Table \ref{tab:tuningbeta}.
Analysis procedures for the static quark potential
and extraction of $r_0/a$
parallel those in Ref. \cite{nf2}.
The simulation parameters and results in this work
are shown in Table \ref{tab:newr0a}.
To avoid finite size effects,
we followed a criterion \cite{Necco}
that the parameter $\beta$ and $L/a$ are 
chosen such that $L/r_0 \sim 3.3$.
Following the above reference,
the number of over-relaxation sweeps
are taken to satisfy $N_{\rm OR} \sim 1.5 (r_0/a)$.

 \section{Continuum extrapolation}
  \label{sec:CutOff}

  \subsection{Step scaling function}
   \label{subsec:SSF}

In this subsection we investigate the cut off dependence of the SSF and
perform the continuum extrapolation 
     \begin{equation}
       \sigma(2,u)
       =
       \lim_{a/L \rightarrow 0}
       \Sigma(2,u,a/L).
     \end{equation}

\begin{figure}[]
   \begin{center}
    \begin{tabular}{cc}
       \scalebox{1.1}{\includegraphics{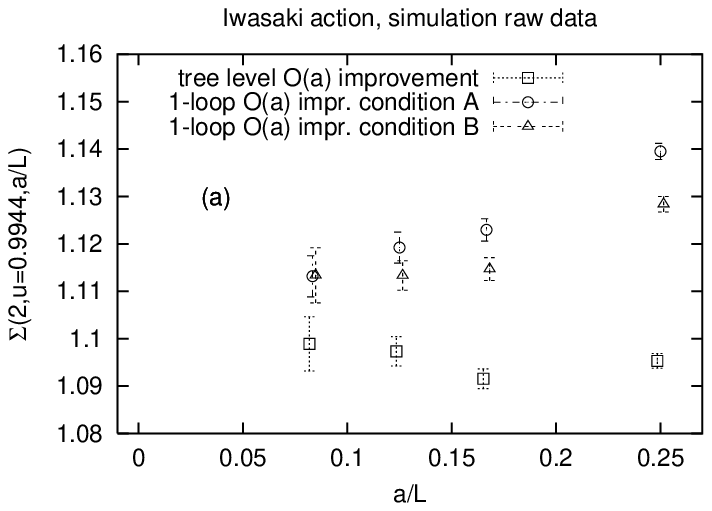}}
       &
       \scalebox{1.1}{\includegraphics{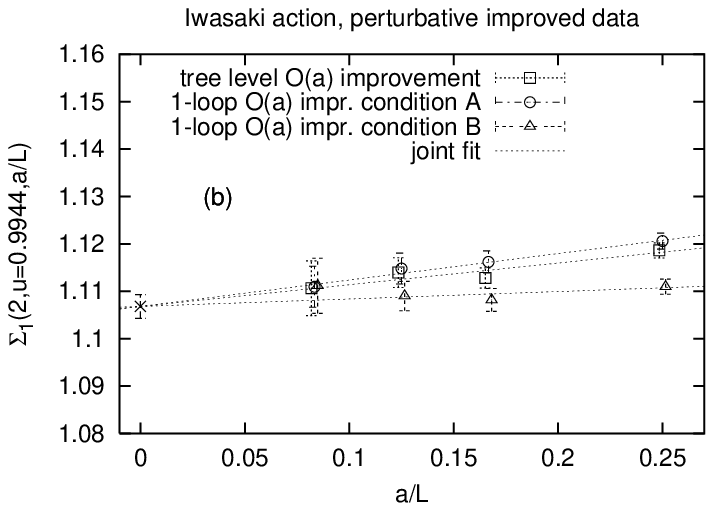}}
       \\
       \scalebox{1.1}{\includegraphics{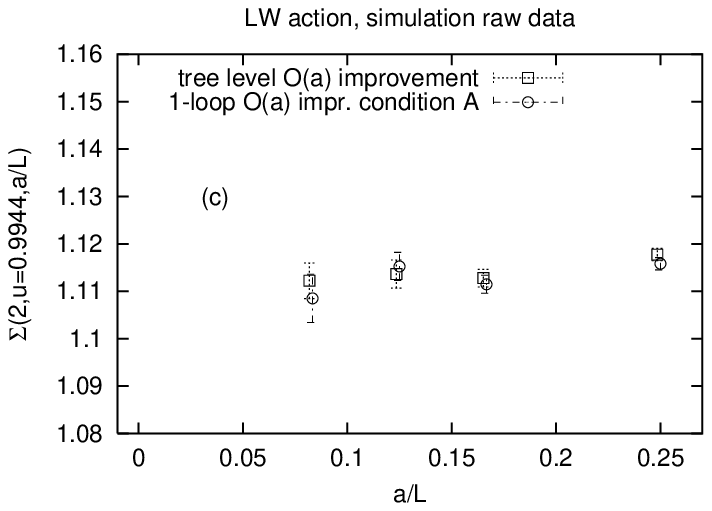}}
       &
       \scalebox{1.1}{\includegraphics{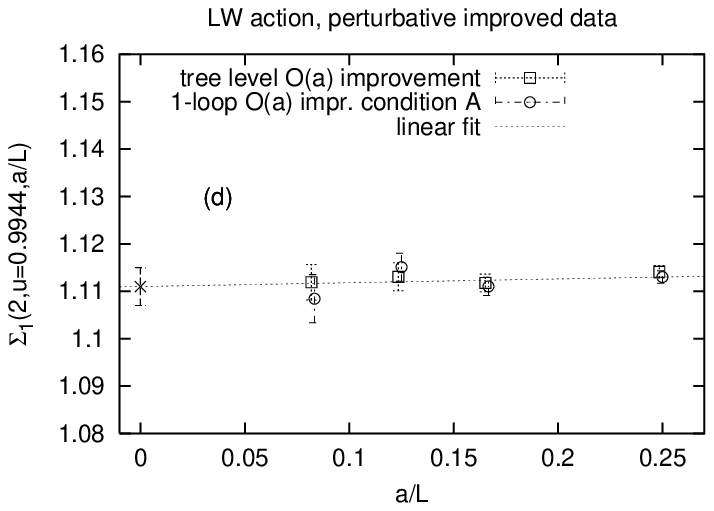}}
       \\
    \end{tabular}
   \end{center}
 \caption{Results of SSF at the weak coupling $u=0.9944$
with the Iwasaki action (upper)
and LW action (lower).
The simulation raw data (left-hand side)
and the data with ``perturbative removal of the lattice artifacts''
(right-hand side) are shown.
The points at $a/L=0$ in (b) and (d)
represent the extrapolated values obtained by
the constrained fit, that the dotted lines indicate the
fitting function, for the Iwasaki action
and by the linear fit, whose fitting function is shown as
dotted line, for
the data of 1-loop O($a$) improved action for the LW action
respectively.}
\label{fig:weak}
\end{figure}

The lattice SSF $\Sigma(2,u,a/L)$ as a function of $a/L$ 
at the weak coupling $u=0.9944$
is shown in Fig. \ref{fig:weak} (a) and (c) for
the Iwasaki action and LW action, respectively.
For the Iwasaki action, even after the one-loop O($a$) improvement
with either the condition A or B,
the scaling violation is still rather large, 
which makes the extrapolation to the 
continuum limit difficult.
To improve the scaling behaviour of the SSF,
we apply a perturbative removal of the lattice artifacts suggested in 
Ref.\cite{su2poly} given by 
\begin{equation}
  \Sigma^{(k)}_1(2,u,a/L)
  =
  \frac{\Sigma^{(k)}(2,u,a/L)}
       {1 + \delta^{(k)}_1(a/L) u},
\end{equation}
where $\Sigma^{(k)}(2,u,a/L)$ is the SSF (simulation raw data)
with the ``$k$''-level O($a$)
improvement coefficient ({\it e.g.,} 
$k=0$: tree level O($a$) improvement case, 
$k=1A$: one-loop O($a$) improvement with condition A case, {\it etc}.).
$\delta^{(k)}_1(a/L)$ is the one-loop relative deviation, given by 
     \begin{eqnarray}
       \delta^{(k)} (2,u,a/L) 
       &=& 
         \frac{\Sigma^{(k)}(2,u,a/L) 
       - \sigma(2,u)}{\sigma(2,u)}
       \nonumber \\ 
       &=& 
       \delta^{(k)}_1(2,a/L) u  
     + O(au^2),
       \label{eqn:devi}
     \end{eqnarray}
whose numerical values are given in Table \ref{tab:deviation}.
This method eliminates not only $O(a)$ but also $O(a^n)$ with $n > 1$
lattice artifacts at one-loop order.
Figure \ref{fig:weak} (b) shows the cut off dependence of $\Sigma_{1}^{(k)}
(2,u,a/L)$.
Indeed the scaling violations are much reduced 
by this method, so that
we can reliably 
take the continuum extrapolation linearly in $a$ as
\begin{equation}
  \Sigma^{(k)}_{1}(2,u,a/L) = \sigma^{(k)}(2,u) + \omega^{(k)}_1(u) a/L,
 \label{eqn:linear}
\end{equation}
where $\sigma^{(k)}(2,u)$ and $\omega^{(k)}_1(u)$ are fit parameters.
In Table \ref{tab:continuumSSF}
we quote the extrapolated value for the Iwasaki action,
which is obtained by a simultaneous fit for $k=0$, $1A$ and $1B$ data
with the constraint that they agree in the continuum limit 
$\sigma^{(k)}(2,u)=\sigma(2,u)$.
The fit has a good $\chi^2/N_{\rm dof}$ 
as listed in Table \ref{tab:continuumSSF}.

As shown in Fig. \ref{fig:weak} (c),
the scaling violations are quite small for the LW action.
This is consistent with the fact found in Ref.\cite{takeda} that 
the lattice artifacts are quite small at one loop.
Moreover, 
the difference between the tree level and one-loop O($a$) improvement
is invisible in this precision,
as a result of the smallness of the
improvement coefficients $c^{P(1)}_{\rm t}$ and $c^{R(1)}_{\rm t}$.
As shown in Fig. \ref{fig:weak} (d),
the perturbative removal of lattice artifacts has almost no effect
except for $L/a=4$, since $\delta^{(k)}_1(2,a/L)$
with $L/a=6$, $8$, $12$ is quite small.
In Table \ref{tab:continuumSSF}
we quote the extrapolated value for the LW action,
obtained by a linear fit to data of the one-loop O($a$) improved action
with the perturbative removal of lattice artifacts. 
For comparison  results of ALPHA Collaboration are also 
included\cite{NPRquench}.

\begin{figure}[]
   \begin{center}
    \begin{tabular}{cc}
       \scalebox{1.1}{\includegraphics{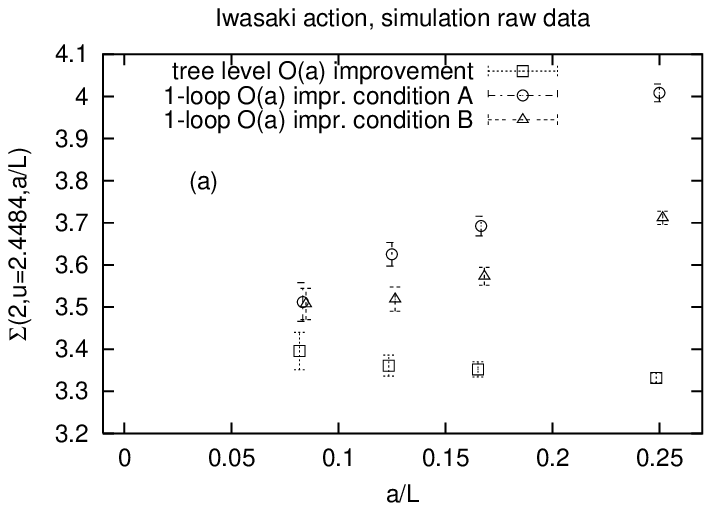}}
       &
       \scalebox{1.1}{\includegraphics{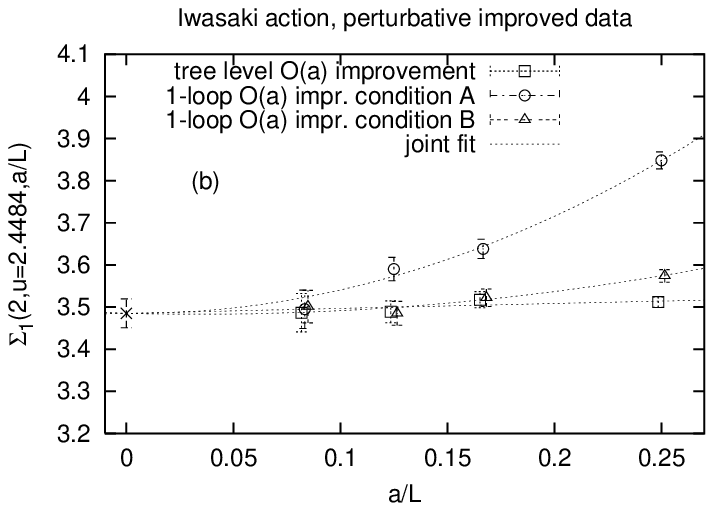}}
       \\
       \scalebox{1.1}{\includegraphics{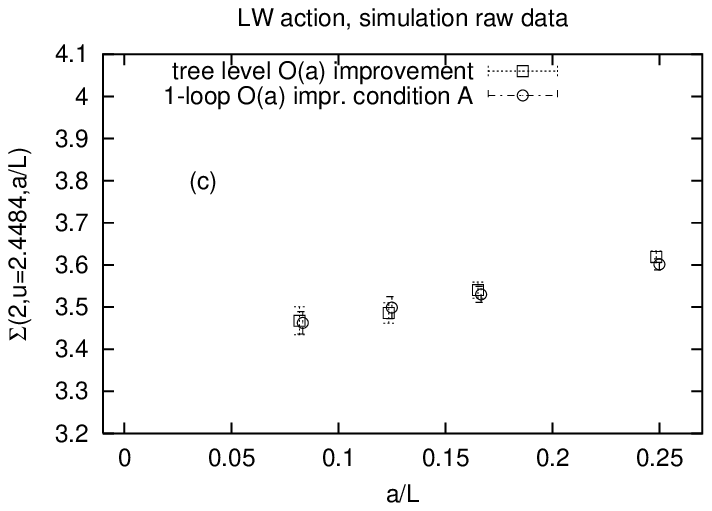}}
       &
       \scalebox{1.1}{\includegraphics{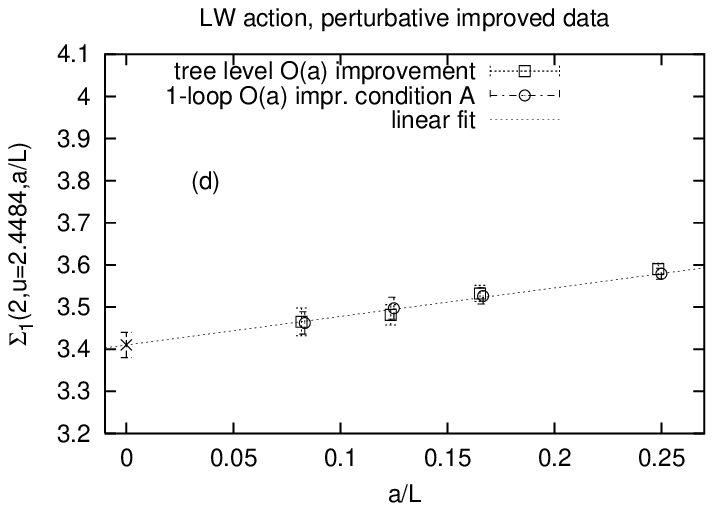}}
       \\
    \end{tabular}
   \end{center}
 \caption{Results of SSF at the strong coupling $u=2.4484$
with the various gauge actions.
Concerning the symbols and the lines,
the same explanation as the weak coupling case is followed.}
\label{fig:strong}
\end{figure}

Results at the strong coupling $u=2.4484$ 
are plotted in Fig. \ref{fig:strong}.
For the Iwasaki action, 
the one-loop O($a$) improvement shows
large lattice artifacts for both conditions A and B, 
particularly for the coarse lattice
(see Fig. \ref{fig:strong} (a)).
As shown in Fig. \ref{fig:strong} (b),
the perturbative removal of lattice artifacts 
well reduces the scaling violation in the case of the condition B, 
but it still remains rather large for the condition A.
Therefore, 
we include a quadratic term in the fitting form, 
\begin{equation}
  \Sigma^{(k)}_1(2,u,a/L) 
= \sigma^{(k)}(2,u) + \omega^{(k)}_1(u) a/L + \omega^{(k)}_2(u) (a/L)^2,
 \label{eqn:linearquadratic}
\end{equation}
for the data of $k=1A$ and $1B$, while
we use the linear fitting form eq. (\ref{eqn:linear}) for
the data of $k=0$.
The extrapolated values obtained with the constraint 
$\sigma^{(k)}(2,u)=\sigma(2,u)$ for a unique continuum value is 
listed in Table \ref{tab:continuumSSF}.
We note that $|\omega^{(1A)}_2(u)/\omega^{(1A)}_1(u)| \approx O(10)$
in the fit for the condition A.  
This suggests that the condition A
accidentally enhances the coefficient of O($a^2$) term.
It does not necessarily mean, however, 
that the one-loop O($a$) improvement itself
is inefficient at this coupling constant.
Indeed the one-loop O($a$) improvement with the condition B
shows good scaling behavior.
As for the LW action,
Fig. \ref{fig:strong} (c) and (d) show that
neither the one-loop O($a$) improvement or 
the perturbative removal works effectively.
Concerning the extrapolation 
we simply use the same procedure as in the weak coupling case,
i.e. the linear fitting form to the perturbative removal data
for one-loop O($a$) improvement.
The result is given in Table \ref{tab:continuumSSF}. 

We observe in Table \ref{tab:continuumSSF}
that the three values obtained with the Iwasaki and LW action in the present work 
and that of ALPHA Collaboration \cite{NPRquench}
at the weak coupling are consistent within 1$\sigma$.  
At the strong coupling, the value for the LW action undershoots relative to the 
others by 1.5--2$\sigma$.
We think that the latter disagreement is caused by a large lattice 
artifact for the LW action, which makes
the choice of the fitting form difficult.
For example, if we assume that $O(a)$ errors for the LW action are negligible,
we can obtain a result consistent with the values of the other actions 
within 1$\sigma$, by using a purely quadratic fitting form. 
Further investigation are needed to clarify this point.

  \subsection{Low energy scale ratio}
    \label{subsec:lowenergyscales}

We now combine $L_{\rm max}/a$ and $a/r_0$ for the Iwasaki action 
to form the ratio $L_{\max}/r_0$, 
and extrapolate it to the continuum limit:
  \begin{equation}
    L_{\rm{max}}/r_0
    =
    \lim_{a/{L_{\rm{max}}} \rightarrow 0}
    (L_{\rm{max}}/a)
    \times
    (a/r_0).
   \label{eqn:contlowenergy2}
  \end{equation}

In the fourth column of Table \ref{tab:scale},
we give the first factor, which is taken from Table \ref{tab:tuningbeta}; 
the error of $L_{\rm max}/a$ is estimated by
propagating that of $\bar{g}^2_{\rm SF}(L_{\max})$.

The second factor $r_0/a$ is given in the third column of Table \ref{tab:scale}. 
This is obtained by an interpolation of the results for $r_0/a$ 
in Table \ref{tab:datar0a} using a polynomial \cite{Wittig}
      \begin{equation}
         \ln (a / r_0)
         =
         c_1 + c_2 (\beta-3) + c_3 (\beta-3)^2 .
         \label{eqn:fitformr0}
      \end{equation}
The fit, plotted in Fig. 4, 
gives
      \begin{eqnarray}
         c_1
         &=&
         -2.193    (6),
         \mbox{\hspace{4mm}}   
         c_2 = -1.344(7),
         \mbox{\hspace{4mm}}   
         c_3 = 0.191(24),
      \end{eqnarray}
with $\chi^2/N_{\rm dof} = 4.10/6$ in the range $2.456 \leq \beta \leq 3.53$.
The error of $r_0/a$ in Table \ref{tab:scale} includes both statistical 
and systematic ones.
We take the central value of $r_0/a$ from the result of the fit above, 
and estimate the systematic error from the difference of $r_0/a$ between
the central value and the result of another fit including 
a $c_4 (\beta-3)^3$ term. 

The combination of the two factors for various $\beta$ values 
are listed in the fifth column in Table \ref{tab:scale}.
The $\beta$ dependence of $L_{\max}/r_0$
can be considered as lattice cut off effects.
For an extrapolation to the continuum limit,
we use a fit form
      \begin{equation}
        a/r_0
        =
        b_1 a/L_{\max}
      + b_2 (a/L_{\max})^2
      + b_3 (a/L_{\max})^3,
       \label{eq:lowfit}
      \end{equation}
where $b_i(i=1,2,3)$ are fit parameters and $b_1$
is the continuum value of the low energy scale ratio, 
rather than fitting $L_{\max}/r_0$ as a function of $a/r_0$.  
In this way one can avoid the correlation of errors which complicates 
the latter fit\footnote{
Since the error on $L_{\max}/a$ is small,
one may perform a continuum extrapolation
of $L_{\max}/r_0$ as a function of $a/L_{\max}$
neglecting the error on the x-axis. 
We observe consistency between the two methods.}.
We apply eq.(\ref{eq:lowfit}) to three sets of data, {\it i.e.,}
data for the tree level O($a$) improvement and those 
of the one-loop O($a$) improvement with the conditions A and B.
For the first set of data, we set $b_3=0$ (a linear fit) and 
exclude the point at $a/L=1/4$.  An alternative fit including that point 
and allowing a non-zero $b_3$ yields a consistent value for $b_1$ within 
errors. 
However, we think that the fit is not so reliable 
since $\chi^2/N_{\rm dof}$ is too small and the linear terms are rather 
large. Therefore we exclude the point and use the linear fit for the remaining
3 points.
In Fig. 5 $L_{\max}/r_0$ is plotted as a function of $a/L_{\max}$. 
Dashed lines are the fit curves eq.(IV.9) divided by $a/L_{\max}$, and 
the point at $a/L_{\max}=0$ shows the extrapolated value $b_1$.

The extrapolated value is given in Table \ref{tab:continuumLr},
together with 
the previous result for the standard Wilson plaquette action\cite{Necco}.
While rather large lattice artifacts are observed
for both standard Wilson plaquette action
and Iwasaki action,
the extrapolated values agree within errors.

 \section{Conclusions and discussions}
  \label{sec:Conclusions}

In this paper, 
we have calculated
the step scaling function (SSF) at the weak and the strong couplings 
for both Iwasaki and LW actions
with the one-loop O($a$) improved 
as well as the tree level O($a$) improved boundary terms.
We have also calculated
the low energy scale ratio for the Iwasaki action
with both tree level and one-loop O($a$) improvements.
The extrapolated values of the SSF at the weak and strong coupling
for various gauge actions
are consistent within 1$\sigma$ and 2.3$\sigma$, respectively.
The low energy scale ratio is also 
consistent between the Iwasaki and plaquette actions within 1$\sigma$.
In conclusion, we have confirmed the universality of both quantities.

We have investigated lattice cut off effects in some detail.
In the extrapolation procedure,
the perturbative removal of lattice artifacts reduces 
the scaling violation of the SSF
for the Iwasaki action
with the tree level O($a$) improvement and the
one-loop O($a$) improvement with the condition B.
Indeed, at the strong coupling at the coarsest lattice $L/a=4$,
cut off effects are of order $1$\% and $3$\%, respectively,
if one compares the extrapolated value
obtained by the constrained fit.
At the weak coupling, they are roughly $1$\% for both cases.
We conclude that for the Iwasaki gauge action
these combinations of improvements are
the good choice for controlling lattice artifacts.
This conclusion is also supported by the fact that
an individual extrapolation to the continuum limit
with the linear fitting form for the data set with 
the tree level O($a$) improvement or the
one-loop O($a$) improvement with the condition B
gives a result consistent with the extrapolated value
estimated from the constrained fit within errors,
at both weak and strong couplings.

As mentioned in the introduction,
this work is the second step
toward $N_{\rm f}=2$ and $3$ simulations.
The present study shows that we should use
the tree level O($a$) improved action
or the one-loop O($a$) improved action with the condition B
in future simulations with dynamical quarks.

\begin{acknowledgements}
This work is supported
in part by the Grant-in-Aid of the Ministry of Education
(Nos. 
13135204, 
13640260, 
14046202, 
14740173, 
15204015, 
15540251, 
15540279, 
15740134, 
16028201, 
16540228, 
16740147, 
16$\cdot$11968). 
S.T. is supported by the JSPS Research Fellowship.
\end{acknowledgements}

\begin{figure}[]
 \begin{center}
     \scalebox{1.2}{\includegraphics{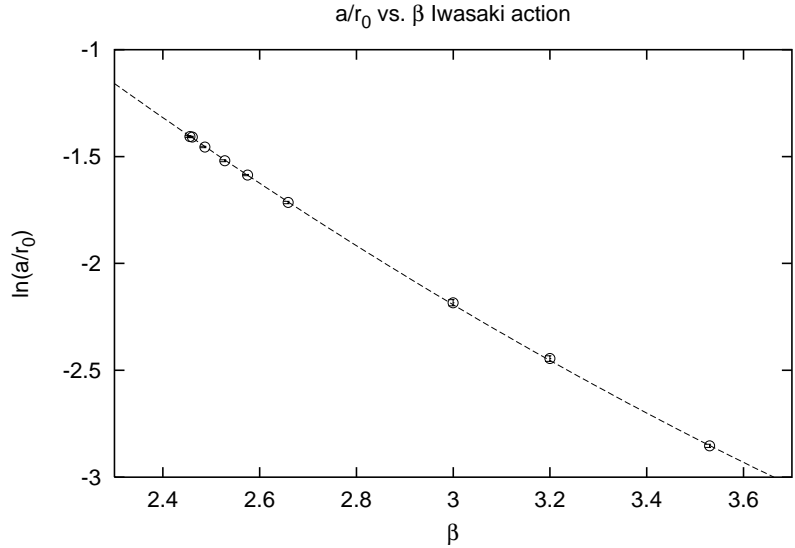}}
      \caption{Interpolation of $r_0/a$ with
phenomenological representation (eq. (\ref{eqn:fitformr0}))
in the range $2.456 \leq \beta \leq 3.53$.}
 \end{center}
 \label{fig:interpolation}
\end{figure}

\begin{figure}[]
 \begin{center}
     \scalebox{1.5}{\includegraphics{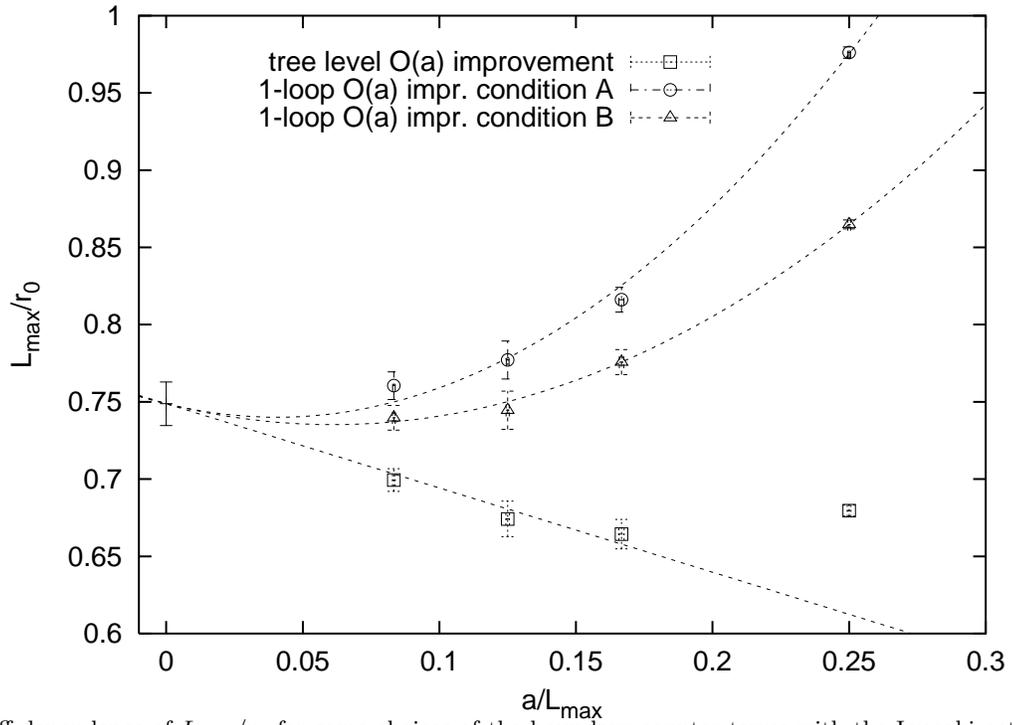}}
      \caption{Cut off dependence of $L_{\max}/r_0$
for some choices of the boundary counter terms with the Iwasaki action.
The error on x-axis are invisible in this scale.
The constrained fit is
represented by dotted lines.
In the fit the point at $a/L_{\max}=1/4$ of the tree level O($a$) 
improvement are not included.
}
 \end{center}
 \label{fig:cutofflowenergyscaleratio}
\end{figure}

\newpage

\begin{table}[ht]
   \begin{center}
\begin{tabular}{cr@{.}lcr@{.}l}
 action & \multicolumn{2}{c}{$c^{P(1)}_{\rm{t}}$ with condition A} 
                            &$c^{P(1)}_{\rm{t}}$ with condition B & \multicolumn{2}{c}{$A_1$}   \\
\hline
Iwasaki &   $0$&$1518$   & $0.04161$  &  $0$&$3036$  \\
LW      &  $-0$&$00297$  &            & $-0$&$00594$  \\
\end{tabular} 
    \caption{The values of $c^{P(1)}_{\rm{t}}$ and $A_1$ 
for the improved gauge actions with each condition.} 
    \label{tab:ct}
\end{center}
\end{table}

\begin{table}[ht]
   \begin{center}
   \begin{tabular}{ccrrlll}	
      action    
   &  degree of O($a$) impr.
   &  \multicolumn{1}{c}{$L/a$}  
   &  \multicolumn{1}{c}{$\beta$}  
   &  \multicolumn{1}{c}{$\bar{g}_{\rm{SF}}^2(L)$}
   &  \multicolumn{1}{c}{$\bar{g}_{\rm{SF}}^2(2L)$}
   &  \multicolumn{1}{c}{$\Sigma(2,u,a/L)$}
   \\ \hline
   & 
   &  $4$ 
   &  $6.5447$
   &  $0.9944(5)$
   &  $1.0953(14)$
   &  $1.0953(16)$
   \\ 
   & 
   &  $6$
   &  $6.8485$
   &  $0.9944(8)$
   &  $1.0915(19)$
   &  $1.0915(21)$
   \\ 
   & \raisebox{1.5ex}[0pt]{tree}
   &  $8$
   &  $7.0733$
   &  $0.9944(12)$
   &  $1.0973(27)$
   &  $1.0973(31)$
   \\ 
   & 
   &  $12$
   &  $7.3765$
   &  $0.9944(18)$
   &  $1.0989(53)$
   &  $1.0989(57)$
   \\ \cline{2-7}
   & 
   &  $4$ 
   &  $6.1467$
   &  $0.9944(5)$
   &  $1.1395(16)$
   &  $1.1395(17)$
   \\ 
   &  one-loop
   &  $6$
   &  $6.5930$
   &  $0.9944(8)$
   &  $1.1230(21)$
   &  $1.1230(24)$
   \\ 
\raisebox{1.5ex}[0pt]{Iwasaki}
   &  condition A
   &  $8$
   &  $6.8799$
   &  $0.9944(13)$
   &  $1.1192(29)$
   &  $1.1192(33)$
   \\ 
   & 
   &  $12$
   &  $7.2547$
   &  $0.9944(14)$
   &  $1.1132(40)$
   &  $1.1132(43)$
   \\ \cline{2-7}
   & 
   &  $4$ 
   &  $6.2258$
   &  $0.9944(5)$
   &  $1.1284(15)$
   &  $1.1284(16)$
   \\ 
   &  one-loop
   &  $6$
   &  $6.6358$
   &  $0.9944(9)$
   &  $1.1147(22)$
   &  $1.1147(24)$
   \\ 
   &  condition B
   &  $8$
   &  $6.9010$
   &  $0.9944(9)$
   &  $1.1133(30)$
   &  $1.1133(31)$
   \\ 
   & 
   &  $12$
   &  $7.2722$
   &  $0.9944(15)$
   &  $1.1134(56)$
   &  $1.1134(58)$
   \\ \hline
   & 
   &  $4$ 
   &  $8.2189$
   &  $0.9944(4)$
   &  $1.1177(12)$
   &  $1.1177(13)$
   \\ 
   &  
   &  $6$
   &  $8.5889$
   &  $0.9944(7)$
   &  $1.1128(17)$
   &  $1.1128(19)$
   \\ 
   & \raisebox{1.5ex}[0pt]{tree}
   &  $8$
   &  $8.8479$
   &  $0.9944(10)$
   &  $1.1136(27)$
   &  $1.1136(30)$
   \\ 
   & 
   &  $12$
   &  $9.2017$
   &  $0.9944(16)$
   &  $1.1122(32)$
   &  $1.1122(37)$
   \\ \cline{2-7}
\raisebox{1.5ex}[0pt]{LW}
   & 
   &  $4$ 
   &  $8.2199$
   &  $0.9944(4)$
   &  $1.1158(12)$
   &  $1.1158(13)$
   \\ 
   &  one-loop
   &  $6$
   &  $8.5957$
   &  $0.9944(7)$
   &  $1.1115(17)$
   &  $1.1115(19)$
   \\ 
   &  condition A
   &  $8$
   &  $8.8406$
   &  $0.9944(10)$
   &  $1.1153(27)$
   &  $1.1153(30)$
   \\ 
   & 
   &  $12$
   &  $9.2060$
   &  $0.9944(16)$
   &  $1.1085(47)$
   &  $1.1085(51)$
    \end{tabular}
    \caption{Results of SSF at the weak coupling $u=0.9944$.}
    \label{tab:weak}
    \end{center}
\end{table}

\begin{table}[ht]
   \begin{center}
   \begin{tabular}{ccrrlll}	
      action    
   &  degree of O($a$) impr.
   &  \multicolumn{1}{c}{$L/a$}  
   &  \multicolumn{1}{c}{$\beta$}  
   &  \multicolumn{1}{c}{$\bar{g}_{\rm{SF}}^2(L)$}
   &  \multicolumn{1}{c}{$\bar{g}_{\rm{SF}}^2(2L)$}
   &  \multicolumn{1}{c}{$\Sigma(2,u,a/L)$}
   \\ \hline
   &
   &  $4$ 
   &  $3.2663$
   &  $2.4484(28)$
   &  $3.332(11)$
   &  $3.332(12)$
   \\ 
   & 
   &  $6$
   &  $3.5754$
   &  $2.4484(52)$
   &  $3.352(16)$
   &  $3.352(18)$
   \\ 
   &\raisebox{1.5ex}[0pt]{tree}
   &  $8$
   &  $3.7872$
   &  $2.4484(40)$
   &  $3.361(24)$
   &  $3.361(25)$
   \\ 
   &
   &  $12$
   &  $4.0996$
   &  $2.4484(65)$
   &  $3.396(43)$
   &  $3.396(44)$
   \\ \cline{2-7}
   &
   &  $4$ 
   &  $2.9628$
   &  $2.4484(29)$
   &  $4.008(21)$
   &  $4.008(21)$
   \\ 
   &  one-loop
   &  $6$
   &  $3.3803$
   &  $2.4484(60)$
   &  $3.692(21)$
   &  $3.692(23)$
   \\ 
\raisebox{1.5ex}[0pt]{Iwasaki}
   &  condition A
   &  $8$
   &  $3.6544$
   &  $2.4484(42)$
   &  $3.625(27)$
   &  $3.625(28)$
   \\ 
   &
   &  $12$
   &  $4.0091$
   &  $2.4484(68)$
   &  $3.512(45)$
   &  $3.512(46)$
   \\ \cline{2-7}
   &  
   &  $4$
   &  $3.0624$
   &  $2.4484(29)$
   &  $3.712(15)$
   &  $3.712(16)$
   \\ 
   &  one-loop
   &  $6$
   &  $3.4395$
   &  $2.4484(56)$
   &  $3.573(20)$
   &  $3.573(21)$
   \\ 
   &  condition B
   &  $8$
   &  $3.6908$
   &  $2.4484(47)$
   &  $3.519(28)$
   &  $3.519(29)$
   \\ 
   &  
   &  $12$
   &  $4.0283$
   &  $2.4484(64)$
   &  $3.518(38)$
   &  $3.518(39)$
   \\ \hline
   &
   &  $4$ 
   &  $4.8992$
   &  $2.4484(24)$
   &  $3.619(13)$
   &  $3.619(14)$
   \\ 
   &
   &  $6$
   &  $5.2786$
   &  $2.4484(49)$
   &  $3.540(18)$
   &  $3.540(20)$
   \\ 
   &\raisebox{1.5ex}[0pt]{tree}
   &  $8$
   &  $5.5325$
   &  $2.4484(42)$
   &  $3.486(23)$
   &  $3.486(24)$
   \\ 
   &
   &  $12$
   &  $5.8878$
   &  $2.4484(62)$
   &  $3.468(32)$
   &  $3.468(33)$
   \\ \cline{2-7}
\raisebox{1.5ex}[0pt]{LW}   &
   &  $4$ 
   &  $4.9055$
   &  $2.4484(25)$
   &  $3.601(13)$
   &  $3.601(13)$
   \\ 
   &  one-loop
   &  $6$
   &  $5.2784$
   &  $2.4484(51)$
   &  $3.530(18)$
   &  $3.530(19)$
   \\ 
   &  condition A
   &  $8$
   &  $5.5332$
   &  $2.4484(40)$
   &  $3.499(26)$
   &  $3.499(26)$
   \\ 
   &
   &  $12$
   &  $5.8867$
   &  $2.4484(58)$
   &  $3.463(25)$
   &  $3.463(27)$
    \end{tabular}
    \caption{Results of SSF at the strong coupling $u=2.4484$.} 
    \label{tab:strong}
    \end{center}
\end{table}

\begin{table}[ht]
   \begin{center}
   \begin{tabular}{crrl}	
      degree of O($a$) impr.
   &  \multicolumn{1}{c}{$L_{\max}/a$}  
   &  \multicolumn{1}{c}{$\beta$}  
   &  \multicolumn{1}{c}{$\bar{g}_{\rm{SF}}^2(L_{\max})$}
   \\ \hline
   &  $4$
   &  $2.7000$ 
   &  $3.480(6)$
   \\
   &  $6$
   &  $3.0057$
   &  $3.480(11)$
   \\
\raisebox{1.5ex}[0pt]{tree}
   &  $8$
   &  $3.2154$
   &  $3.480(11)$
   \\ 
   &  $12$
   &  $3.5219$
   &  $3.480(13)$
   \\ \hline
   &  $4$
   &  $2.4594$ 
   &  $3.480(7)$
   \\ 
      one-loop
   &  $6$
   &  $2.8556$
   &  $3.480(14)$
   \\ 
      condition A
   &  $8$
   &  $3.1047$
   &  $3.480(11)$
   \\ 
   &  $12$
   &  $3.4496$
   &  $3.480(16)$
   \\ \hline
   &  $4$
   &  $2.5382$ 
   &  $3.480(6)$
   \\ 
      one-loop
   &  $6$
   &  $2.8921$
   &  $3.480(11)$
   \\ 
      condition B
   &  $8$
   &  $3.1376$
   &  $3.480(12)$
   \\ 
   &  $12$
   &  $3.4734$
   &  $3.480(14)$
    \end{tabular}
    \caption{Tuning of $\beta$
             at $u=3.480$ for Iwasaki action.}
    \label{tab:tuningbeta}
    \end{center}
\end{table}

\begin{table}[ht]
   \begin{center}
   \begin{tabular}{ccrcr}
      $\beta$
   &  $(L/a)^4$
   &  \multicolumn{1}{c}{$r_0/a$}
   &  $N_{\rm OR}$
   &  \multicolumn{1}{c}{$N_{\rm conf}$}
   \\ \hline
      $3.00$
   &  $32^4$ 
   &  $8.88(13)$
   &  $15$
   &  $400$
   \\
      $3.53$
   &  $56^4$ 
   &  $17.35(13)$
   &  $24$
   &  $88$
    \end{tabular}
    \caption{Simulation parameters and results performed
in this work for $r_0/a$ with
the Iwasaki action. 
$N_{\rm OR}$ and $N_{\rm conf}$ 
indicate the number of over-relaxation sweep and configuration
respectively.}
    \label{tab:newr0a}
    \end{center}
\end{table}

\begin{table}[ht]
   \begin{center}
   \begin{tabular}{cr@{.}lc}	
      $\beta$
   &  \multicolumn{2}{c}{$r_0/a$}
   &  reference
   \\ \hline
      $2.456$
   &  $4$&$080(16)$ 
   &  \cite{nf2}
   \\ 
      $2.461$
   &  $4$&$089(14)$ 
   &  \cite{topology}
   \\ 
      $2.487$
   &  $4$&$286(15)$ 
   &  \cite{nf2}
   \\ 
      $2.528$
   &  $4$&$570(21)$ 
   &  \cite{nf2}
   \\ 
      $2.575$
   &  $4$&$887(16)$ 
   &  \cite{nf2}
   \\ 
      $2.659$
   &  $5$&$556(30)$ 
   &  \cite{topology}
   \\ 
      $3.000$
   &  $8$&$88(13)$ 
   &  in this work
   \\ 
      $3.200$
   &  $11$&$53(15)$ 
   &  \cite{okamoto}
   \\ 
      $3.530$
   &  $17$&$35(13)$ 
   &  in this work
    \end{tabular}
    \caption{$\beta$ versus $r_0/a$ with
the Iwasaki action. The value of $r_0/a$ are taken from
the references quoted in the last column.}
    \label{tab:datar0a}
    \end{center}
\end{table}

\begin{table}[ht]
   \begin{center}
     \begin{tabular}{cccccc} 
       & \multicolumn{3}{c}{Iwasaki action} 
       & \multicolumn{2}{c}{LW action} 
       \\ \hline
       & & & & & \\
       \raisebox{1.5ex}[0pt]{$L/a$}  
     & \raisebox{1.5ex}[0pt]{$\delta^{(0)}_1$} 
     & \raisebox{1.5ex}[0pt]{$\delta^{(1A)}_1$} 
     & \raisebox{1.5ex}[0pt]{$\delta^{(1B)}_1$} 
     & \raisebox{1.5ex}[0pt]{$\delta^{(0)}_1$} 
     & \raisebox{1.5ex}[0pt]{$\delta^{(1A)}_1$} 
       \\ \hline
   4& -0.02096 & 0.01700 & 0.01577 & 0.003278 & 0.002536  \\
   6& -0.01922 & 0.00608 & 0.00592 & 0.000911 & 0.000417  \\
   8& -0.01499 & 0.00399 & 0.00395 & 0.000527 & 0.000156  \\
  12& -0.01064 & 0.00201 & 0.00200 & 0.000296 & 0.000049  \\
    \end{tabular}
    \caption{1-loop deviations for various gauge actions 
and with $k$-level O($a$) improvement.}
    \label{tab:deviation}
    \end{center}
\end{table}

\begin{table}
\begin{center}
\begin{tabular}{lcccc}
action     & 
$\sigma(2,u=0.9944)$ & 
$\chi^2/N_{\rm dof}$ & 
$\sigma(2,u=2.4484)$ & 
$\chi^2/N_{\rm dof}$ \\
\hline
Iwasaki   & 1.107(3)  & 1.5/8 & 3.485(34) & 1.9/6 \\
LW        & 1.111(4)  & 2.0/2 & 3.410(30) & 0.1/2 \\
plaquette \cite{NPRquench}& 1.110(11) & 3.3/2 & 3.464(40) & 0.9/4
\end{tabular}
    \caption{The extrapolated values of the SSF with various gauge actions
for both the weak and strong coupling.
The values in the case of the plaquette action
are quoted from the reference as indicated.}
    \label{tab:continuumSSF}
\end{center}
\end{table}

\begin{table}[]
   \begin{center}
   \begin{tabular}{crr@{.}lr@{.}lll}	
      degree of O($a$) impr.
   &  \multicolumn{1}{c}{$\beta$}  
   &  \multicolumn{2}{c}{$r_0/a$}  
   &  \multicolumn{2}{c}{$L_{\rm{max}}/a$}  
   &  \multicolumn{1}{c}{$L_{\rm{max}}/r_0$}
   \\ \hline
   &  $2.7000$ 
   &  $5$&$886(26)$
   &  $4$&$000(11)$
   &  $0.680(4)$
   \\ 
   &  $3.0057$
   &  $9$&$03(12)$
   &  $6$&$000(35)$
   &  $0.664(10)$
   \\ 
\raisebox{1.5ex}[0pt]{tree}
   &  $3.2154$
   &  $11$&$87(19)$
   &  $8$&$000(41)$
   &  $0.674(12)$
   \\ 
   &  $3.5219$
   &  $17$&$16(14)$
   &  $12$&$00(8)$
   &  $0.699(7)$
   \\ \hline
   &  $2.4594$ 
   &  $4$&$098(12)$
   &  $4$&$000(9)$
   &  $0.976(4)$
   \\ 
      one-loop
   &  $2.8556$
   &  $7$&$352(59)$
   &  $6$&$000(33)$
   &  $0.816(8)$
   \\ 
      condition A
   &  $3.1047$
   &  $10$&$29(16)$
   &  $8$&$000(31)$
   &  $0.777(12)$
   \\ 
   &  $3.4496$
   &  $15$&$78(16)$
   &  $12$&$00(7)$
   &  $0.761(9)$
   \\ \hline
   &  $2.5382$ 
   &  $4$&$625(12)$
   &  $4$&$000(8)$
   &  $0.865(3)$
   \\ 
      one-loop
   &  $2.8921$
   &  $7$&$735(73)$
   &  $6$&$000(28)$
   &  $0.776(8)$
   \\ 
      condition B
   &  $3.1376$
   &  $10$&$74(17)$
   &  $8$&$000(38)$
   &  $0.745(12)$
   \\ 
   &  $3.4734$
   &  $16$&$22(15)$
   &  $12$&$00(7)$
   &  $0.740(8)$
    \end{tabular}
    \caption{The low energy scale at various values of $\beta$ for Iwasaki action.}
    \label{tab:scale}
    \end{center}
\end{table}

\begin{table}
\begin{center}
\begin{tabular}{lcc}
action     &  
$L_{\max}/r_0$ & 
$\chi^2/N_{\rm dof}$ \\
\hline
Iwasaki               & 0.749(14) & 3.93/ 5 \\
plaquette \cite{Necco}& 0.738(16) &
\end{tabular}
    \caption{The extrapolated values of the low energy scale
ratio with various gauge actions.
The values in the case of the plaquette action
are quoted from the reference as indicated.}
    \label{tab:continuumLr}
\end{center}
\end{table}

\appendix


\end{document}